# GPU-MetaD: Full-Life-Cycle GPU Accelerated Metadynamics with Machine Learning Potentials

Haoting Zhang,[1] Qiuhan Jia,[1] Zhennan Zhang, [1] Yijie Zhu,[1] Zhongwei Zhang,[1] Junjie Wang,[1,†] Jiuyang Shi [1,†] Zheyong Fan[2] and Jian Sun[1,†]

[1] *National Laboratory of Solid State Microstructures, School of Physics and Collaborative Innovation Center of Advanced Microstructures, Nanjing University, Nanjing 210093, China.*

[2] *College of Physical Science and Technology, Bohai University, Jinzhou, P. R. China.*

## ABSTRACT

Large-scale molecular dynamics simulations with high accuracy have been increasingly popular for their capability to bridge the gap between atomistic modeling and mesoscale phenomena. Both machine learning potentials and enhanced sampling approaches offer substantial improvements in high-accuracy simulation efficiency, which can be further boosted through GPU acceleration. However, an efficient framework combining these advances for extending simulations to large systems and long timescales remains elusive. In this work, we proposed a full-life-cycle GPU accelerated metadynamics simulations package GPU-MetaD. Benchmarking across molecular, interface, and bulk systems demonstrates that GPU-MetaD efficiently handles diverse atomic systems and delivers an order-of-magnitude performance improvement. Building on this demonstrated capability, it enables ab-initio-level rare-event sampling for systems comprising millions of atoms on a typical single GPU. This capability allows us to reveal a previously unknown size-dependent two-step nucleation mechanism in gallium nitride (GaN), highlighting the potential of GPU-MetaD for uncovering complex rare events in realistic large-scale materials systems.

## INTRODUCTION

To investigate more realistic systems and complex problems in the fields of physics, biology and material science, there is a growing demand of combining large-scale modeling with high-accuracy interatomic potentials in molecular dynamics (MD) simulations for studies such as microsecond protein folding [1], device-scale atomistic modeling [2], and multi-million-atom nucleation processes [3].Recent advances in two key directions have made such simulations increasingly achievable. First, the emergence of data-driven machine learning potentials (MLPs) has enabled ab-initio accuracy at a fraction of the computational cost [4–7]. Second, the continuous development of graphics processing units (GPUs) has dramatically increased the parallel efficiency of MD simulations, giving rise to a new generation of GPU-based simulation engines capable of handling systems of unprecedented scale [8–11]. These two developments have significantly extended the accessible spatial and temporal scales of MD simulations, thereby narrowing the gap between atomistic simulations and experimentally accessible conditions. On the other side, enhanced sampling techniques, such as umbrella sampling (US) [12], replica exchange MD (REMD) [13], Gaussian accelerated MD (GaMD) [14] and metadynamics (MetaD) [15,16], have become indispensable for overcoming the intrinsic timescale limitations of MD, enabling efficient reconstruction of free-energy surfaces. Among these methods, metadynamics has proven versatile, owing to its adaptive biasing scheme and ease of integration with existing MD software frameworks.

---

*Corresponding author. J.S (jiansun@nju.edu.cn); J.W. (wangjunjie@nju.edu.cn); J.Shi. (sjy@smail.nju.edu.cn)

Building upon the methodological and computational advances, several enhanced sampling frameworks and packages have been proposed [17,18] and successfully applied to a variety of frontier problems [19–22]. However, when extending simulations to larger and more complex systems, the existing frameworks are still limited in both integration and efficiency, which constrains their scalability and practical applicability in large scale problems. To address this gap, we developed GPU-MetaD, a fully GPU-accelerated metadynamics simulation framework built upon the high-performance GPUMD engine [8,23]. GPU-MetaD provides an efficient and user-friendly environment that integrates enhanced sampling algorithms with GPUMD and neuroevolution potential (NEP) [5,8].

The accuracy, robustness, and scalability of GPU-MetaD were validated through extensive tests on molecular, interfacial, and bulk systems, offering mechanistic insights into protein folding, water dissociation at oxide surfaces, and phase transitions in semiconductors respectively. Our results show that GPU-MetaD achieves over an order of magnitude speedup compared with existing CPU-based implementations, while preserving excellent numerical consistency. By integrating enhanced sampling algorithms with GPUMD, GPU-MetaD provides an efficient and scalable platform for high-accuracy simulations of realistic systems. This capability opens new opportunities for predictive modeling of complex chemical reactions and materials transformations, thereby promoting mechanistic understanding and guiding the design of biomacromolecules, catalytic processes, and next-generation materials.

# RESULTS AND DISCUSSION

## Frameworks of GPU-MetaD

A brief workflow of GPU-MetaD is shown in Fig. 1. Building on the GPUMD engine, we developed an end-to-end enhanced sampling workflow entirely implemented on GPUs. GPU-MetaD is developed based on PyTorch [24], a widely-used GPU computing framework with an extensive user base and open-source accessibility. A typical enhanced sampling script is illustrated in Fig.S1. Users can employ our predefined workflow templates, reference our predefined CVs and biases, and flexibly combine them using Python scripts. Additionally, users can also customize CVs and load into our GPUMD-GPU-MetaD interface, offering out-of-the-box functionality for different enhanced sampling workflows.

Once the script is loaded into the GPUMD workflow, the GPUMD-GPU-MetaD interface automatically extracts necessary data from the molecular dynamics process to perform enhanced sampling calculations. Numerical computations of CVs and corresponding bias-driven forces are computed in parallel using PyTorch tensor computation framework and automatic differentiation module. We designed an input-output flows based on the GPUMD runtime processes to ensure the alignment of CV computation outputs with GPUMD outputs, named GPU-Sampling trajectory analysis (GSTA) module, facilitating post-processing and data analysis.

## Alanine dipeptide torsion

For decades, alanine dipeptide has served as a canonical model in computational biophysics, owing to its well-characterized $\Phi$-$\Psi$free energy landscape. This simple yet informative system enables systematic evaluation of enhanced sampling methods and provides a conceptual link between small peptide dynamics and protein folding behavior [16,25–28]. To validate our full-life-cycle GPU workflow, we employed datasets sampled in the Amberff19SB [29] as the ground truth force field according to previous studies [30], and sampled trajectories were subsequently used to construct a NEP model, with training details provided in Fig. S2.

In alanine dipeptide, the two backbone torsional angles annotated in Fig. 2a, Φ(C-N-$C_\alpha$-C) and Ψ(N-$C_\alpha$-C-N), serve as the CVs as a typical scheme. We performed a 20 ns NVT metadynamics simulation by setting $\sigma = 0.35$ rad, $W_k = 1.2$ kJ/mol, and the resulting CV evolution and FES is shown in Fig. 2b and Fig. 2c respectively. Considering the intrinsic error of MLPs, our simulation successfully reproduced multiple metastable basins on the $\Phi-\Psi$ free energy landscape compared with previous works. Specifically, the simulation recovered the well-known minima corresponding to the right-handed α-helix region (Φ ≈ –60°, Ψ ≈ –40°), the β-sheet region (Φ ≈ –120°, Ψ ≈ 120°), as well as additional metastable states such as the left-handed α-helix region (Φ ≈ 60°, Ψ ≈ 60°) [30].

## Water dissociation on rutile (110) surface

Water dissociation, especially on oxide surfaces, has long been a central topic in heterogeneous catalysis [31–34]. In these systems, surface oxygen atoms with dangling bonds can interact with hydrogen atoms from water molecules, thereby facilitating the water-splitting process. Consequently, determining the adsorption energy between surface oxygen and hydrogen is crucial for elucidating the mechanism of surface water dissociation. In recent studies, Chen et al. [34] uncovered facet-dependent, multi-step water dissociation mechanisms on low-index $TiO_2$ surfaces, highlighting the fundamental role of surface structure in governing catalytic water splitting. Among these surfaces, the rutile (110) surface serves as a prototypical model and has been extensively investigated both experimentally and theoretically.

From a publicly available dataset [34], we sampled 1918 rutile (110) surface configurations with adsorbed water molecules and trained a NEP model. The training details are illustrated in Fig. S3. Using the distance between a surface oxygen atom and its nearest hydrogen atom as the CV, we performed a 15 ns NVT well-tempered metadynamics simulation with CV evolution shown in Fig. 3b. Our calculations reconstruct the FES in Fig. 3a and yield an adsorption energy barrier for the surface O–H bond of approximately 21.48 kJ/mol, which agrees well with previous reports within the uncertainty range of enhanced sampling methods [34]. The simulation results highlight the consistency and convenience of our framework in achieving ab initio-level accuracy for surface catalytic processes.

## B4-B1 phase transition in Large-scale systems

For crystalline systems, many dynamical processes exhibit inherently large characteristic length scales. However, molecular dynamics simulations using relatively small supercells often struggle to capture such behaviors, as the imposed periodicity can artificially induce homogeneous behaviors. This so-called size effect hampers realistic simulations of large-scale dynamical phenomena. For example, Yao et al. and related studies reported pronounced size effects in the B4–B1 phase transition of gallium nitride, yet the limited simulation sizes prevented a complete understanding of the transition pathway [21,35–37]. In this work, our framework enables simulations at significantly larger scales, allowing us to uncover previously inaccessible transition pathways.

Based on an open dataset [37], we trained a force field for GaN and performed metadynamics simulations for systems ranging from 27648 to 2.2 million atoms. We selected the coordination number (C.N.) and the triaxial lattice parameters as CVs and ran NPT simulation at 300K and 50 GPa. In the simulation with 27648 atoms, we noticed that the lattice undergoes repeated tensile and compressive deformations, which ultimately give rise to a grid-like network of banded structures composed of 5-fold-coordinated atoms, which is in good agreement with previous results obtained under near conditions [37]. Nuclei consisting of 6-fold-coordinated atoms emerge within the grid-like banded structure. In simulations with 750,000 atoms or fewer, we observed the cooperative nucleation of a tilted strip-like B1 phase. Evidently, this strip touches the periodic boundary, indicating that the system still suffers from finite-size effects. In the 1.3 million atom simulation, we clearly identified a sheet-like B1 nucleus, which continued to grow toward the boundaries and eventually evolved into a strip, leading to the formation of the B1 phase. The simulation with 2.2 million atoms revealed a multi-site nucleation pathway, where several sheet-like B1 nuclei first appeared within the B4 phase. These nuclei gradually expanded and ultimately formed a polycrystalline B1 phase. Fig. 4a presents the simulation results for systems containing 2.2 million atoms which provides clear evidence with the CV evolution shown in Fig. 4b. By increasing the system size, we effectively mitigated finite-size effects and demonstrated the potential of our software package for uncovering novel physical mechanisms.

Beside the general picture described above, the hexagonal-transition path reveals a novel size effect with the two-step nucleation behavior. In Fig. S5 we present states during the hexagonal-transition path, as the B4 (hexagonal diamond) state turning into B1 (NaCl) state during a hexagonal-like (h-MgO) state [21,37]. The trajectory snapshots of the 50 thousand-, 750 thousand- and 2.2 million atoms show in Fig. S6-8 respectively. In the smallest system, the B1 lamellar nuclei within the single shear band remain nearly co-oriented and expand both along and across the shear-band direction, giving rise to an almost single-crystalline B1 domain. In contrast, the 750 thousand- and 2.2 million-atom systems develop a markedly different morphology: the former hosts a single penetrating shear band in Fig. 4b, whereas the latter exhibits multiple intersecting shear bands in Fig. 4c, with more details shown in Fig. 5 and Fig. S10 respectively. Within these bands, two distinct orientations of B1 lamellar nuclei emerge and quickly fills the band as clarified in Fig. 5f and Fig. S9. During this process, the overall band width remains nearly constant as shown in Fig. 5e. This leads to the formation of a polycrystalline B1 band, which subsequently thickens as the systems transforms into a fully polycrystalline B1 phase. As the system size increases, distinct nucleation pathways can be observed, evolving from unidirectional nucleation within a single shear band, to multidirectional nucleation within a single shear band, and ultimately to multidirectional nucleation across multiple shear bands, underscoring the critical importance of large-scale simulations in overcoming finite-size effects and capturing the full complexity of nucleation behavior in extended systems.

## Efficiency of GPU-MetaD package

We evaluated the hardware efficiency of this module in metadynamics simulations across different system sizes. Keeping the NEP model and simulation configuration fixed, we measured the simulation speed on conventional hardware. Across simulations of GaN ranging from 50,000 to 1.3 million atoms, our GPU-MetaD software package consistently demonstrates high computational efficiency, which is only about 2x slower than GPUMD without enhanced sampling. In contrast, the mainstream approach based on the LAMMPS[4] and PLUMED[35] interface is approximately one order of magnitude slower than our method. The interface between GPUMD and PLUMED suffers from performance bottlenecks due to GPU–CPU data transfer overhead and the lack of robust parallel support. As a result, not only is the computation significantly slower, but its performance further degrades as the system size increases. The detailed performance is shown in Fig. 6. All tests applied the same NEP model. Our GPU-MetaD framework was executed on NVIDIA RTX4090. The LAMMPS-PLUMED framework was ran on Intel 9242 CPU with 256G of memory. And the GPUMD-PLUMED framework, which requires both CPU and GPU computation ability, was ran on a workstation with Intel 6230R*2 and NVIDIA RTX5880. It can be asserted that our framework utilizes computational resources with maximum efficiency at a comparable cost, thereby enabling computations of larger systems.

# CONCLUSIONS

In this work, we developed the GPU-MetaD package to provide an intuitive, customizable, and user-friendly enhanced sampling scheme, which takes advantages in computational power and efficiency. Moreover, our approach integrates with GPUMD and NEP, constituting a fully GPU-based metadynamics simulation solution. Users simply need to organize their datasets and select the CVs to perform efficient metadynamics simulations with MLPs on GPUs. We anticipate that GPU-MetaD will pave the way for researchers seeking enhanced sampling simulations especially in large-scale systems, providing a powerful tool for investigating the dynamic mechanisms of crystals and biomacromolecules.

# METHODS

## Metadynamics Method

In conventional MD, the Boltzmann distribution severely hinders access to high-free-energy regions, rendering rare events like barrier-crossing processes unobservable on practical simulation timescales. Bias-based enhanced sampling techniques, such as metadynamics, overcome this limitation by progressively depositing bias potentials that flattens free-energy differences between metastable states and barriers, thereby increasing the transition probabilities [15,16,38].

In a dynamic system with an interaction potential $U_0$, metadynamics works by periodically depositing bias potential $V_b(\mathbf{s};\mathbf{s}_\tau,\tau)$ at the states visited by the system, thus giving the biased interaction:

$$U_b(\mathbf{s}) = U_0(\mathbf{s}) + V_b(\mathbf{s})$$

Where $V_b(\mathbf{s}) = \sum_\tau V_b(\mathbf{s};\mathbf{s}_\tau,\tau)$. $s$ denotes the states of the system, which is usually referred to as the collective variable (CV) and $\tau$ denotes the deposition frame in the simulation, a typical form of the bias potential is the Gaussian-shaped bias:

$$V_b(\mathbf{s}_\tau;\tau) = w_k \exp(-\frac{(\mathbf{s}-\mathbf{s}_\tau)^2}{2\sigma^2})$$

Where $\sigma$ denotes the width of the gaussian hills. Therefore, it can progressively fill the minima of the free-energy surface and thus increase the probability of escaping local minima. In metadynamics procedure, CVs typically assume distinct values corresponding to different local minima and have much fewer degrees of freedom than the full dynamic parameters, thus characterizing system states and slow transition modes near potential barriers. The procedure mostly converges while the FES is flatten, and thus it can be reconstructed by unbiased ensemble possibility:

$$P_{unbias}(\mathbf{s}) = \frac{\left\langle \delta[\mathbf{s}-\mathbf{s}(\mathbf{R})]e^{\beta V_b(\mathbf{s})} \right\rangle_{U_b}}{\left\langle e^{\beta V_b(\mathbf{s})} \right\rangle_{U_b}}$$

$$F(\mathbf{s}) = -\frac{1}{\beta} log\, P_{unbias}(\mathbf{s})$$

where $\langle\cdot\rangle_{U_b}$stands for averaging over the biased ensemble.

## Data availability

The data that support the findings of this study are available from the corresponding author upon reasonable request. All data generated or analysed during this study are included in this published article and its Supplementary Information.

## Code availability

The code used to generate and analyse the data in this study is also included in the Supplementary Information. The code of GPU-MetaD package is available at [GPU-Sampling] (https://github.com/bigd4/GPUMD/tree/GPU-Sampling) and the code for build GPU-MetaD script is available at [pyGAS](https://gitlab.com/tensor4627/pygas).

## ACKNOWLEDGMENTS

We gratefully acknowledge the financial support from the National Natural Science Foundation of China (grant number. 12125404, T2495231, 12504277), the Basic Research Program of Jiangsu (Grant BK20253009, BK20233001, BK20241253), the Jiangsu Funding Program for Excellent Postdoctoral Talent (2024ZB002, 2024ZB075), the Postdoctoral Fellowship Program of CPSF (Grant GZC20240695), the Fundamental and Interdisciplinary Disciplines Breakthrough Plan of the Ministry of Education of China, the AI & AI for Science program of Nanjing University, Artifcial Intelligence and Quantum physics (AIQ) program of Nanjing University, and the Fundamental Research Funds for the Central Universities. The calculations were carried out using supercomputers at the High Performance Computing Center of Collaborative Innovation Center of Advanced Microstructures, the high-performance supercomputing center of Nanjing University.

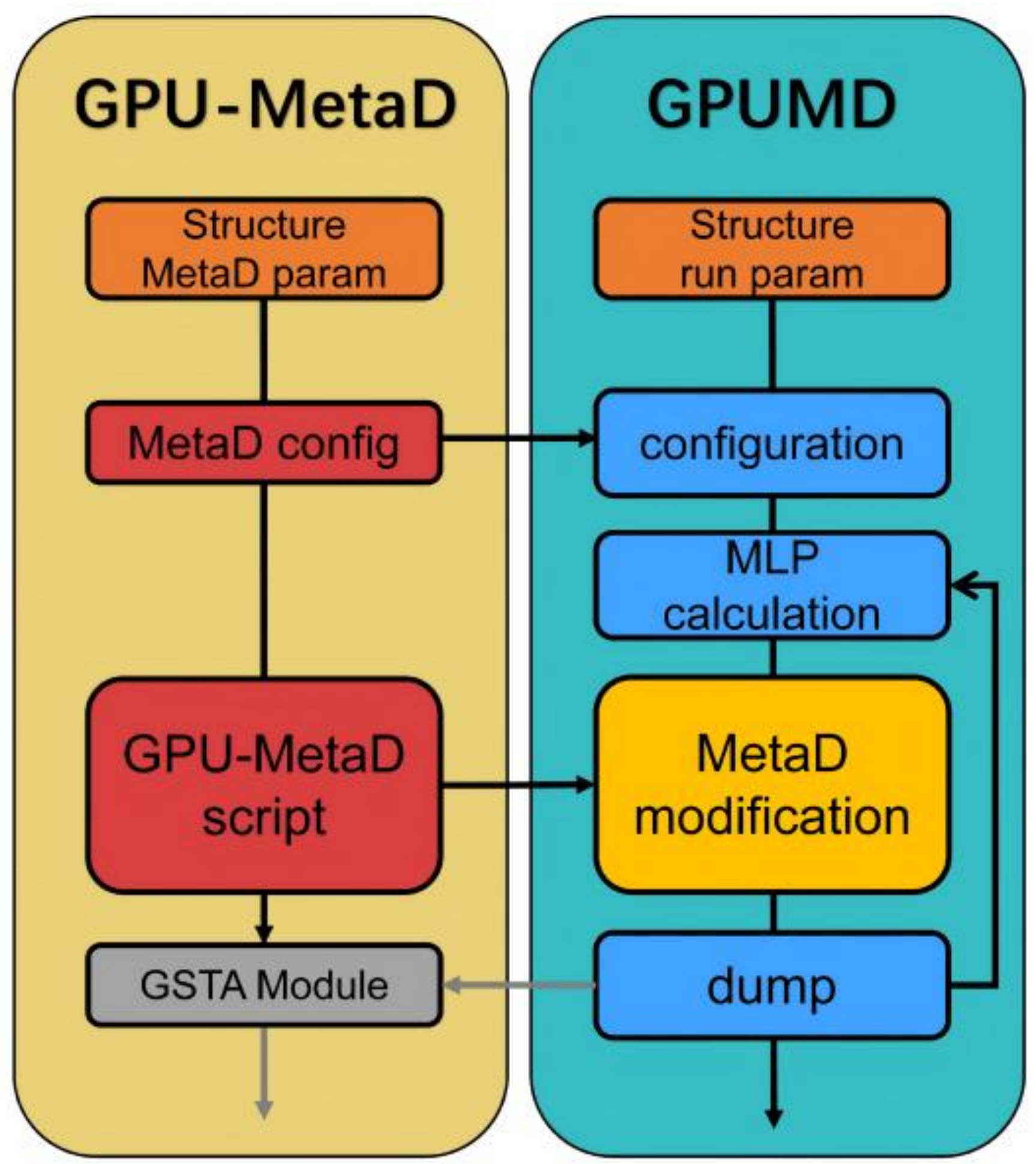

FIG 1. The flow chart of GPU-MetaD and its collaboration pattern with GPUMD. Users can define GPU-MetaD script using a python script with customized configurations. With our GPU-MetaD-interface compiled in GPUMD, GPU-MetaD script can be loaded into GPUMD workflow and automatically modify it into a metadynamics workflow. Simultaneously, the GPU-MetaD script can instantly turns into an analysis tool for any output MD trajectory, which is an optional module independent with GPUMD workflow.

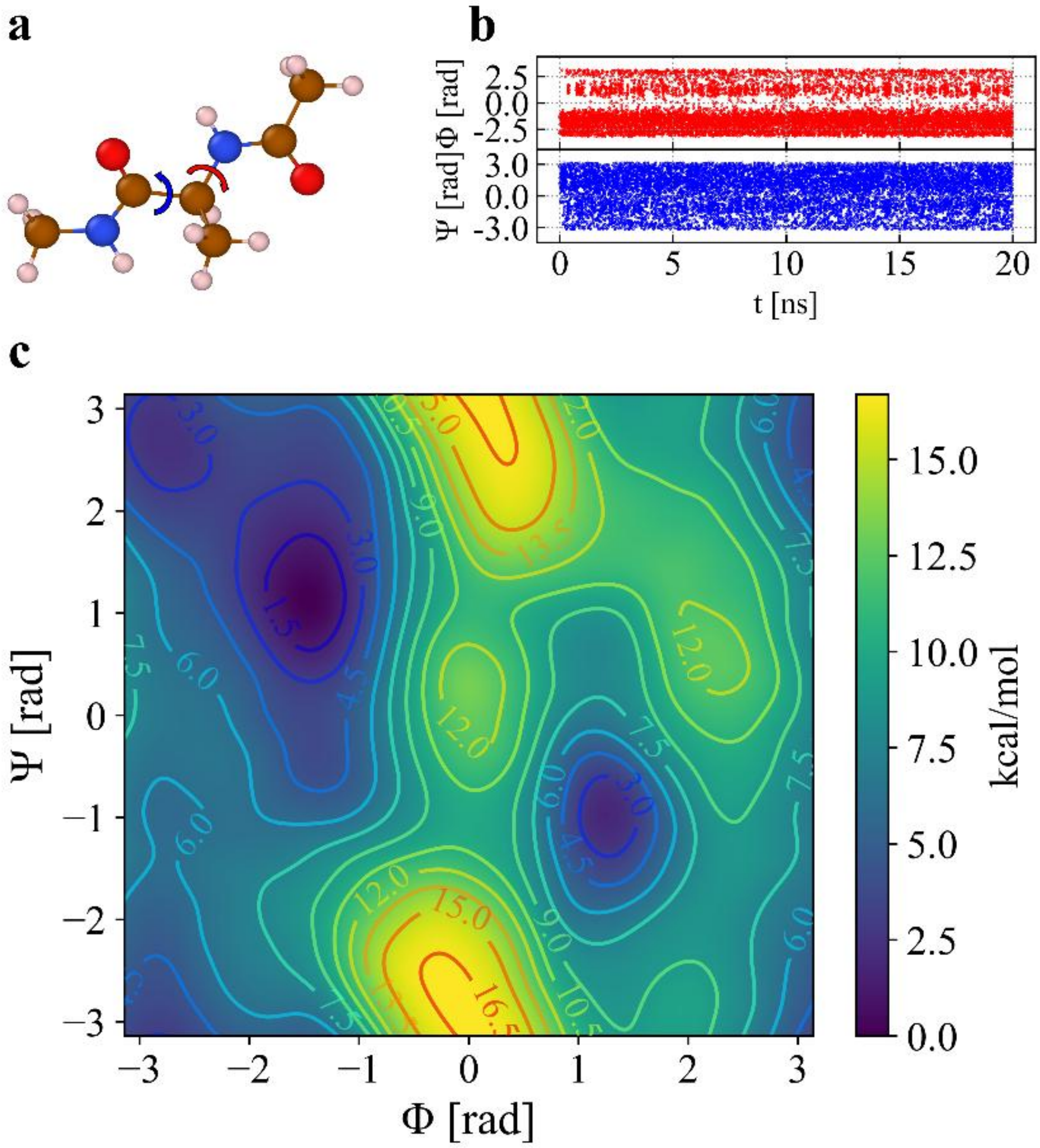

FIG 2. The 20ns NVT metadynamics simulation result of alanine dipeptide. (a) The structure of alanine dipeptide. Red and blue arrows refer to torsion angles (Φ,Ψ) respectively. (b)The evolution of the two backbone torsional angles $\Phi(\text{C-N-C}_\alpha\text{-C})$ and $\Psi(\text{N-C}_\alpha\text{-C-N})$ which serve as the CVs. (c) Metadynamics rebuilt FES with respect to two torsional angles.

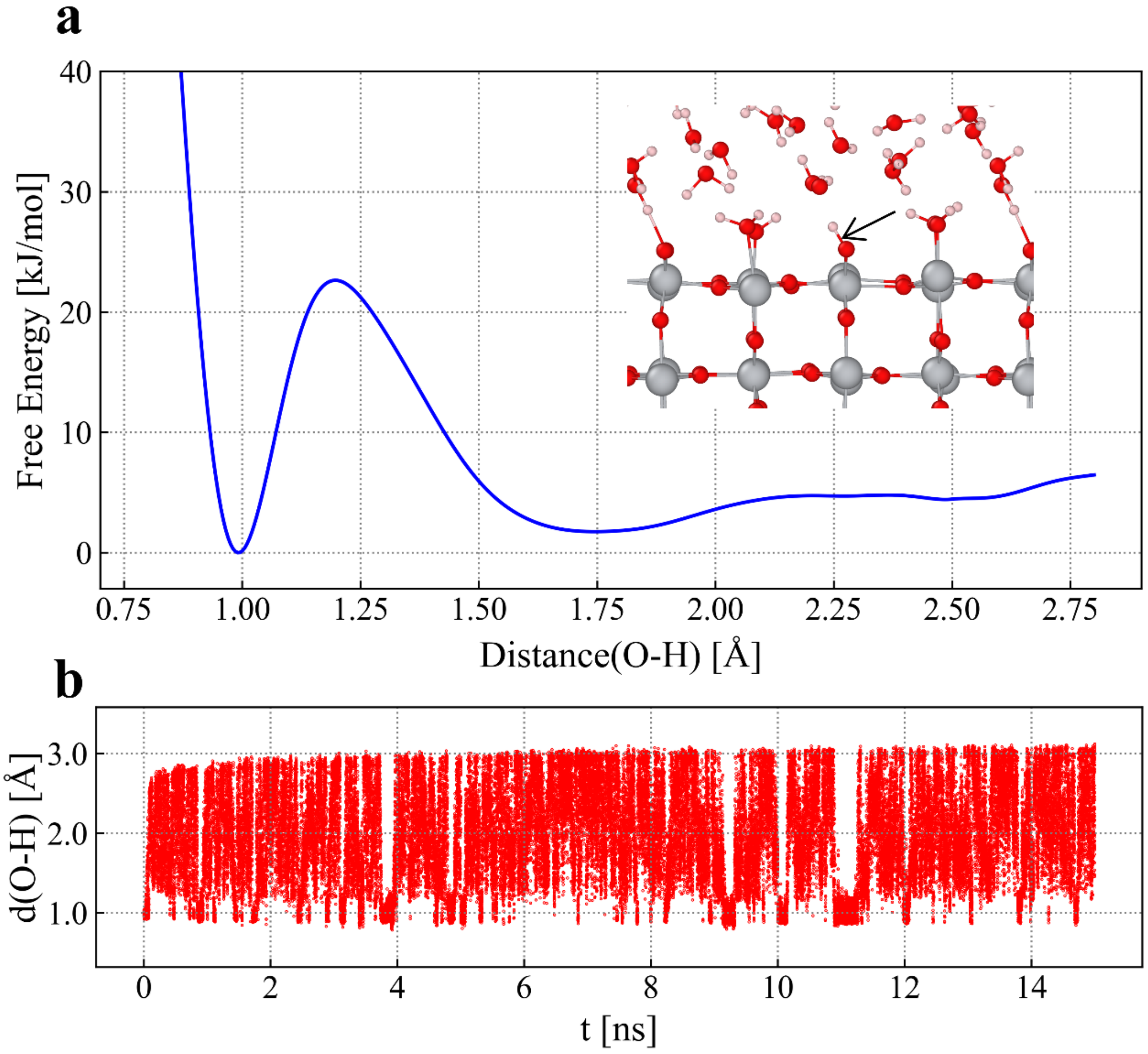

FIG 3. The NVT well-tempered metadynamics simulation result of water-rutile (110) interface. (a)The FES of rutile (110) surface rebuilt by biasing on the bond of surface oxygen and hydrogen in water. The structure shows rutile (110) surface attached with water molecules where the black arrows points to the bond which we set as the CV. (b) The evolution of distance between a surface oxygen atom and its nearest hydrogen during the simulations ) which serves as the CV.

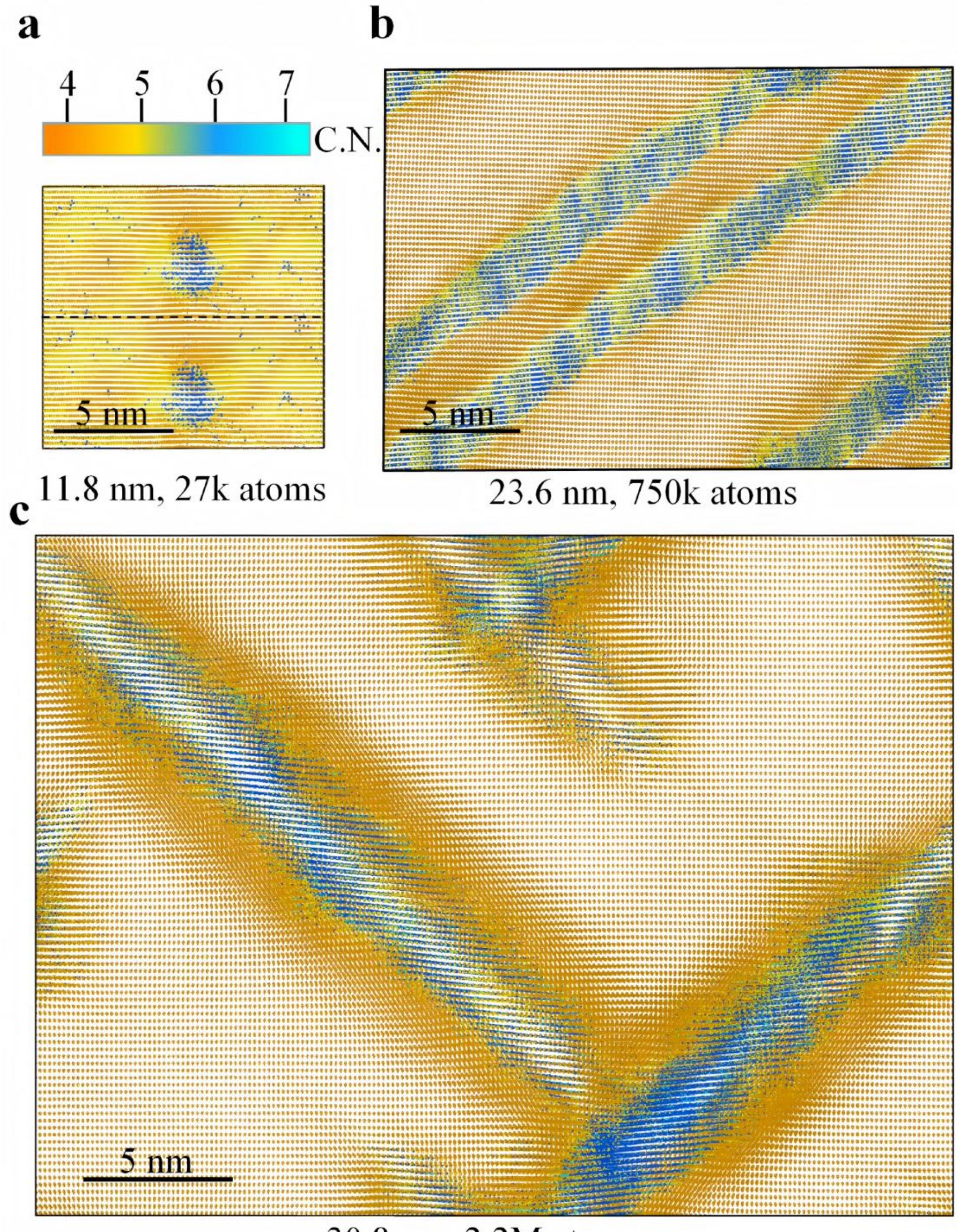

FIG 4. The metadynamics result of GaN B4-B1 phase transition in scales range from 27 thousand to 2.2 million atoms, about 10 to 30 nm, with the length and number of atoms indicated beneath each snapshot, showing diverse transition pathway as size enlarging. (a) The structure with B1 nucleus in B4 phase colored by coordination numbers with 27 thousand atoms, the cell has a 1x2 replication for better vision. (b) The structure with B1 lamellar nuclei in a hexagonal-like shear band with 750 thousand atoms. (c) The structure with B1 lamellar nuclei in multi-directional shear bands with 2.2 million atoms.

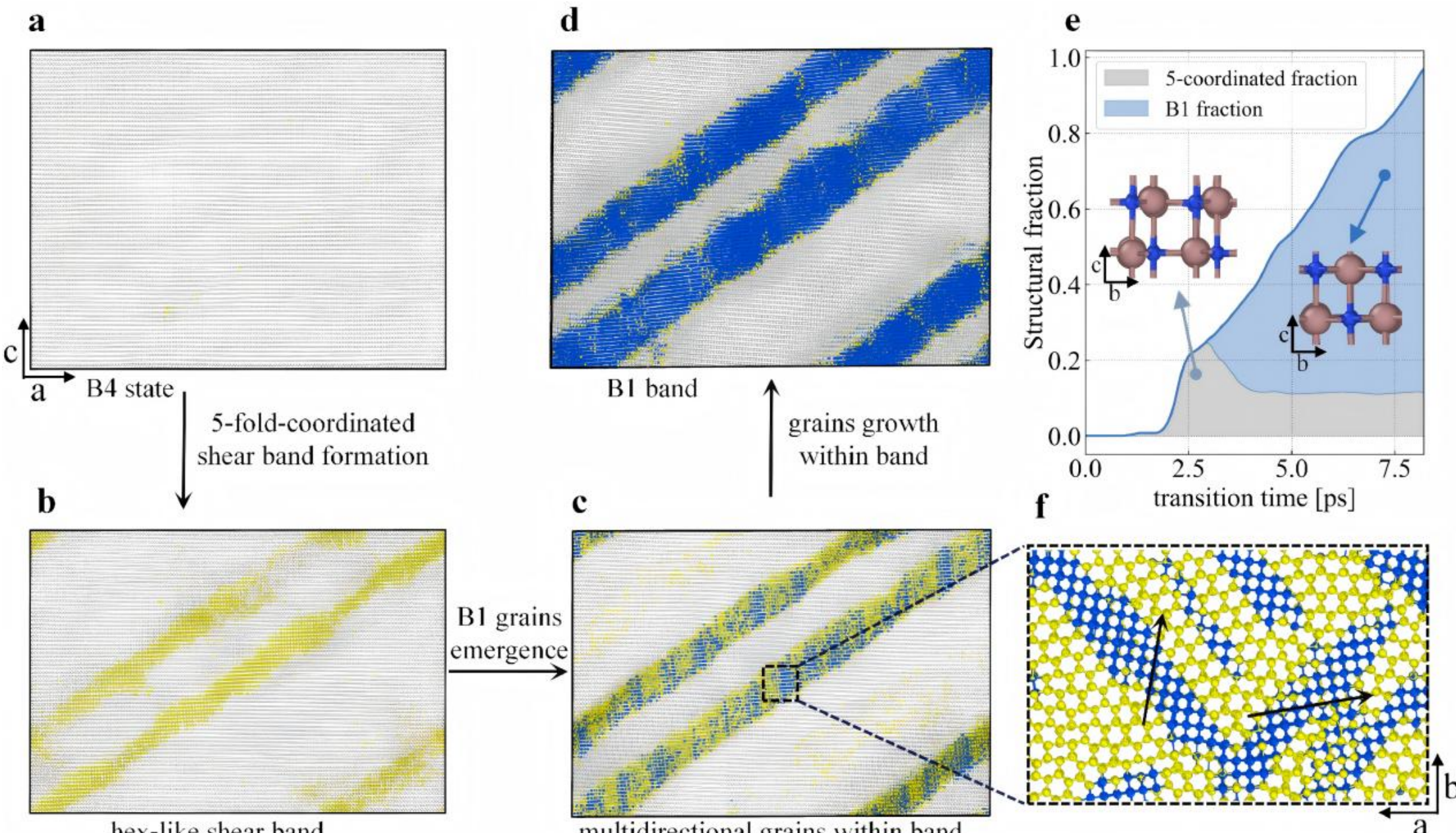

FIG 5. (a-d) The hexagonal pathway of GaN B4-B1 phase transition. The front view of the structure shares the same color code with Fig. 4 except 4-fold-coordinated B4 phase colored by write for better vision. A shear band of 5-fold-coordinated hexagonal-like state forms in the B4 phase and expand slowly. And then blue-colored B1 nuclei emerge and grow in the yellow-colored 5-fold-coordinated region, turning it into a B1 band. (e) The area plot of structural fraction of different states during the transition of 750 thousand atoms of GaN respectively. Structures corresponding to each region are denoted by light and deep blue arrows. the panel indicate that B1 domain fills the shear band quickly as they form, causing the 5-fold-coordinated hex-like region to shrink to a negligible and stable level. Then the B1 band continues to broaden and leading to a polycrystalline final state with 5-fold-coordinated grain boundary. (f) A sliced view of a crystalline shear band. Two distinct orientations of plate-like B1 nuclei are identified within the shear band. Representative nuclei are denoted with arrows indicating their characteristic orientations respectively. These nuclei rapidly grow and expand within the shear band, eventually coalescing to form a polycrystalline B1 phase.

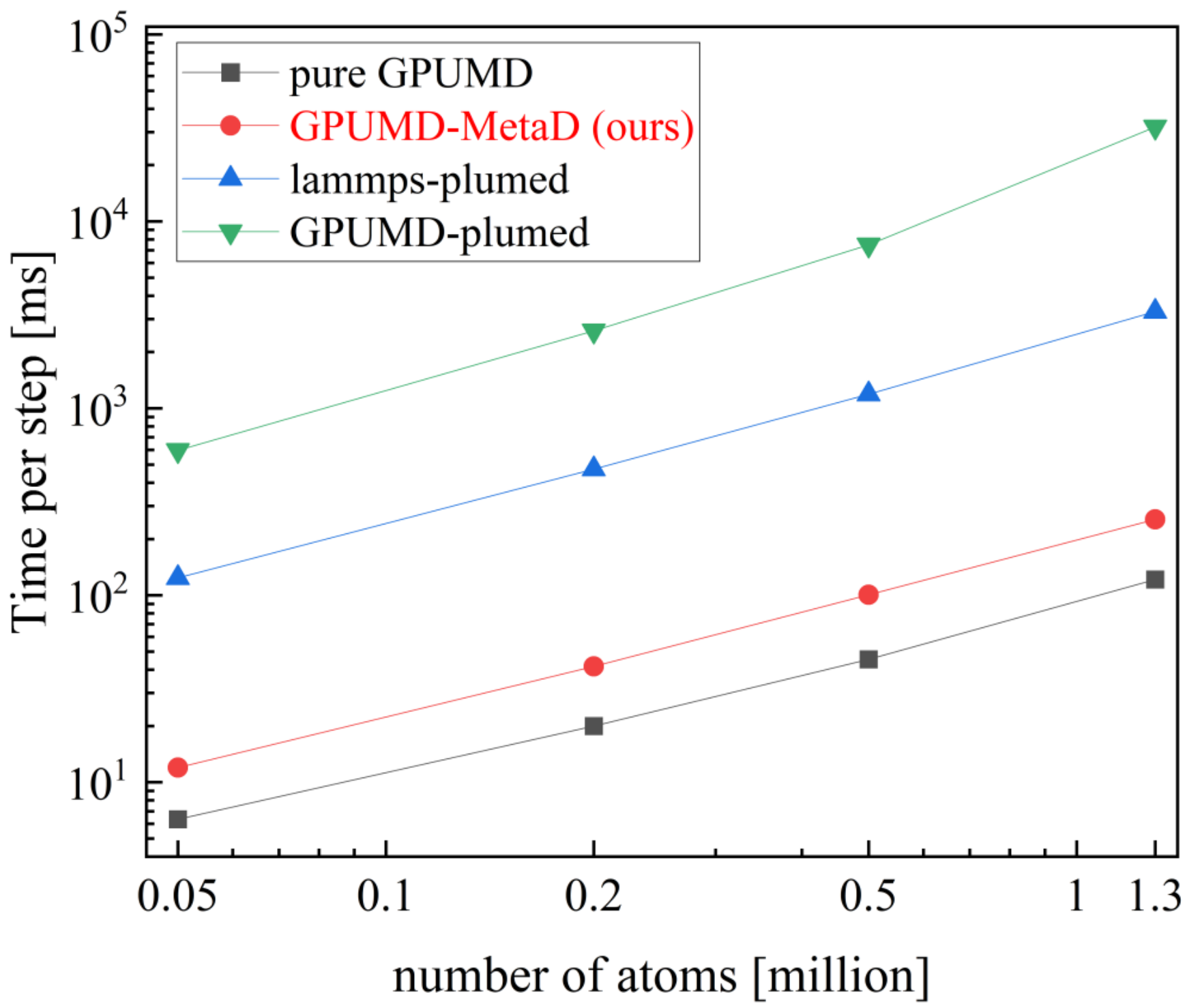

FIG 6. The real time cost of simulations with one simulation steps of different frameworks. The red line represent our GPU-MetaD packages which beats the mainstream LAMMPS-PLUMED metadynamics frameworks with a speed advantages of over 10 times and delivers a performance comparable to that of pure GPUMD.